\documentclass[%
 aapm,
 mph,%
 amsmath,amssymb,
 reprint,%
superscriptaddress
]{revtex4-2}

\usepackage{multirow}

\usepackage{graphicx}% Include figure files
\usepackage{dcolumn}% Align table columns on decimal point
\usepackage{bm}% bold math

\usepackage{xcolor}

\newcommand{\beginsupplement}{%
        \setcounter{table}{0}
        \renewcommand{\thetable}{S\arabic{table}}%
        \setcounter{figure}{0}
        \renewcommand{\thefigure}{S\arabic{figure}}%
     }

\begin{document}

\preprint{AAPM/123-QED}

\title{Wave-driven phase wave patterns in a ring of FitzHugh-Nagumo oscillators}

\author{Daniel Cebrián-Lacasa}
\affiliation{Laboratory of Dynamics in Biological Systems, Department of Cellular and Molecular Medicine, KU Leuven, Herestraat 49, 3000 Leuven, Belgium}
\affiliation{Centre for Engineering Biology, University of Edinburgh, Edinburgh EH9 3BF, United Kingdom}
\author{Marcin Leda}
\affiliation{Centre for Engineering Biology, University of Edinburgh, Edinburgh EH9 3BF, United Kingdom}
\author{Andrew B. Goryachev}
\affiliation{Centre for Engineering Biology, University of Edinburgh, Edinburgh EH9 3BF, United Kingdom}
\author{Lendert Gelens}
\affiliation{Laboratory of Dynamics in Biological Systems, Department of Cellular and Molecular Medicine, KU Leuven, Herestraat 49, 3000 Leuven, Belgium}
\email{lendert.gelens@kuleuven.be}

\date{\today}

\begin{abstract}
We explore a biomimetic model that simulates a cell, with the internal cytoplasm represented by a two-dimensional circular domain and the external cortex by a surrounding ring, both modeled using FitzHugh-Nagumo systems. The external ring is dynamically influenced by a pacemaker-driven wave originating from the internal domain, leading to the emergence of three distinct dynamical states based on the varying strengths of coupling. The range of dynamics observed includes phase patterning, the propagation of phase waves, and interactions between traveling and phase waves. A simplified linear model effectively explains the mechanisms behind the variety of phase patterns observed, providing insights into the complex interplay between a cell's internal and external environments.
\end{abstract}

\keywords{FitzHugh-Nagumo, Phase Waves, Traveling Waves, Phase Patterning, Excitable Media, Relaxation Oscillations}
\maketitle

\section{Introduction} 

The FitzHugh-Nagumo (FHN) model, introduced by Richard FitzHugh in 1961 \cite{FitzHugh1961} and extended spatially by Jinichi Nagumo in 1962 \cite{Nagumo1962}, stands as a seminal activator-inhibitor model in neuroscience. Comprising two polynomial ordinary differential equations (ODEs) – one cubic and one linear – the model is relatively simple:
\begin{equation}
    \begin{array}{ @{} l  @{} }
        u_t=-u^3+u-v, \\[\jot]
        v_t=\varepsilon(u-bv+a).\\
    \end{array}
    \label{EqFHN_ODE}
\end{equation}

The FHN model captures a wide array of dynamical behaviors, including relaxation oscillations, type II excitability, and bistability \cite{Rocsoreanu2000}. Including spatial coupling through diffusion or advection, this set of ODEs transforms into a set of partial differential equations (PDEs), leading to even more complex dynamics such as traveling waves and spatially extended patterns \cite{Nagumo1962,McKean1970,Meron1992,Winfree1991}. Initially conceived to model neuronal activity, the FHN model's versatility has seen its application extend beyond neuroscience, finding utility in fields as diverse as optics, cardiology, and broader biological contexts\cite{cebrianlacasa2024decades}.\\

This study draws inspiration from biological systems, where many regulatory networks are described by activator-inhibitor pairs, which can be modeled by the FHN model. Our primary motivation stems from the complex cell cycle regulatory system, where multiple parts exhibit activator-inhibitor dynamics We specifically focus on how two key regulatory modules collectively coordinate cell division during cytokinesis.\\

On the one hand, interactions between two protein complexes regulate mitosis in the cell cytoplasm: cyclin B - Cdk1, consisting of the protein cyclin B and an enzyme called cyclin-dependent-kinase 1 (Cdk1), and the Anaphase Promoting Complex / Cyclosome (APC/C) \cite{novak1993numerical,pomerening2005systems,gonze2001model,de2021modular,rombouts_DDE_2023}. Additionally, the interplay between Rho GTPase and F-Actin governs cytoskeletal behaviors at the cell cortex \cite{bement2015activator,michaud2022versatile,bement2024patterning}. Actin proteins bind to each other to form filaments, and such filaments, when polymerizing against a cell’s membrane, produce a force that can deform the membrane.
\\

Recent experimental studies in starfish oocytes and embryos have shown that cortical dynamics respond to internal cytoplasmic signals, i.e., cyclin B-Cdk1 enzyme activity\cite{wigbers2021hierarchy, bischof2017cdk1}. By inhibiting one of its positive regulators, Ect2 \cite{su2011targeting}, a cytosolic gradient of cyclin B-Cdk1 controls actin-driven waves of contraction \cite{wigbers2021hierarchy}. However, studies in other model organisms (i.e., frog egg extracts and fly embryos) have found pacemaker-driven waves of Cdk1 activity traveling through the cytoplasm \cite{chang2013mitotic,nolet2020nuclei,puls2024mitotic,deneke2016waves}.  \\

In this work, we characterize the range of dynamical behaviors that can occur at the cell cortex when driven by a cytoplasmic wave traveling at constant speed. We consider this driving to be unidirectional, as no biochemical coupling from the cortex to the cytoplasm has been identified. While membrane deformations could, in principle, influence the cytoplasm as well, recent work demonstrated that under small deformations, these influences are negligible \cite{klughammer2018cytoplasmic}.\\

We utilize a dual FHN system setup to simulate the intricate dynamics of a cell's cytoplasm and cortex. The inner system, representing the cytoplasm, exerts a unidirectional inhibitory influence on the outer system, analogous to the cortex, through pacemaker-driven waves. This setup is designed to reflect the key dynamics of the Cyclin B-Cdk1 (APC/C) and Rho GTP (F-Actin) networks.  By using a system of two coupled FHN models, we focus on the generic properties of wave-driven patterning rather than aiming to describe the detailed biology. As such, this analysis could be relevant for other systems where an activator-inhibitor region is driven by a traveling wave. All simulations are conducted in dimensionless spatial and temporal units to maintain the model's applicability across other contexts. \\ 

The paper is structured as follows: Section \ref{sec:sectionII} outlines the model, its coupling mechanism, and the chosen parameters, aligning with the fundamental properties of the biological networks in question. Detailed exploration of these biochemical interactions, however, is beyond this paper's scope. Section \ref{sec:sectionIII} explores three distinct dynamics within the external ring. Sections \ref{sec:sectionIV} and \ref{sec:sectionV} characterise these dynamics and their origin using simplified settings. Section \ref{sec:sectionVI} illustrates the role of the diffusion rates on these dynamics. In Section \ref{sec:sectionVII}, we then discuss our findings, relating them to existing research and outlining their potential implications. The final section, \ref{sec:sectionVIII}, recapitulates the study's key takeaways and suggests avenues for future research.

\begin{figure*}
    \centering
    \includegraphics[width=\textwidth]{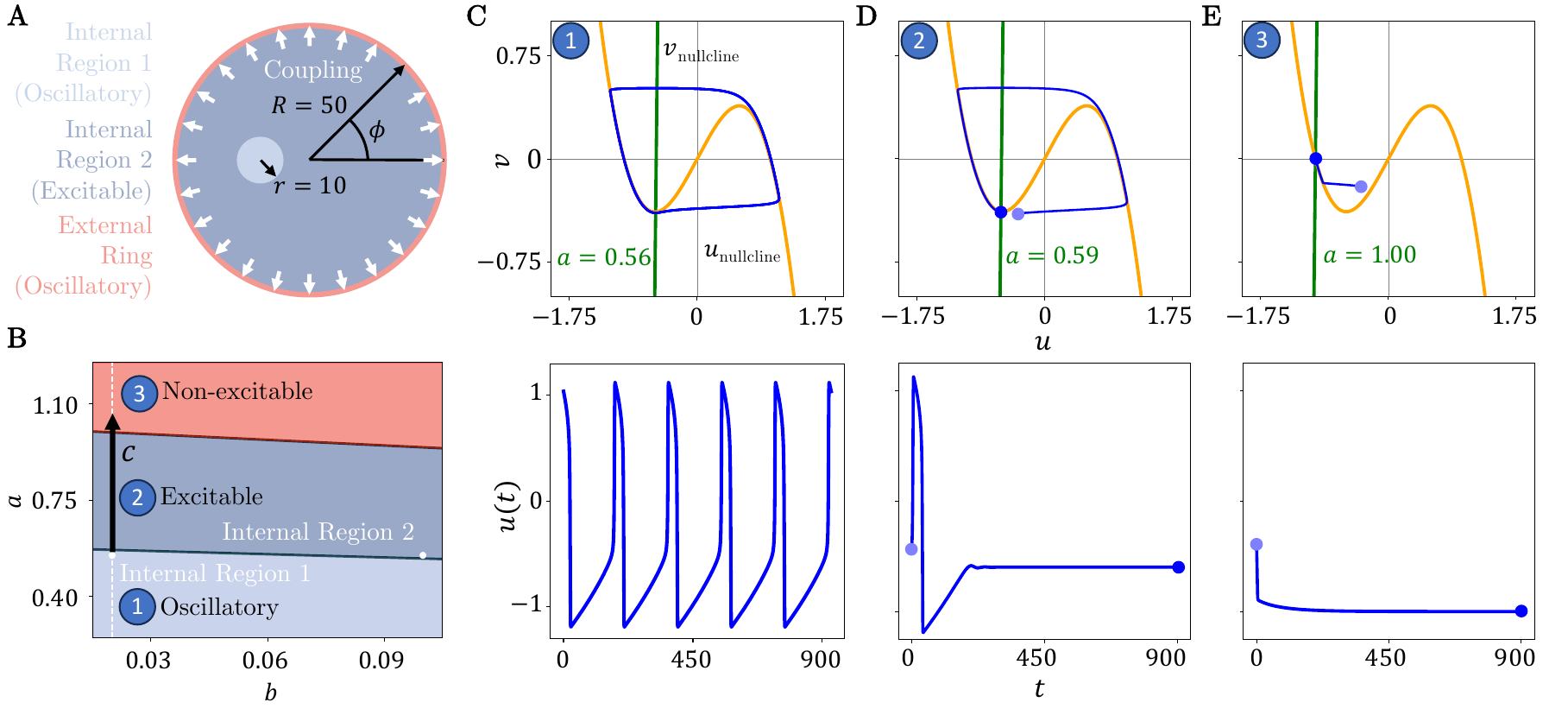}
    \caption{\textbf{Model consisting of two coupled FHN Systems}
A. The depicted system consists of a disc (internal system) encircled by a ring (external system), each governed by the FHN equation within a circular area of radius $R=50$. The internal system is divided into two distinct zones: an oscillatory Region 1 centered within a radius $r=10$ at $\boldsymbol{x}=(-25,0)$; and an adjacent excitable Region 2 (see model parameters at Tab. \ref{tab:modelParameters}). The external system mirrors the parameters of Region 1 but incorporates a coupling factor $c$ that modulates its parameter $a(u_i)$ based on the internal field $u_i$, as illustrated by the white dashed line and the black arrow in panel B. B. Representation in the parameter space ($a,b$) that illustrates all the areas with distinct dynamical behaviors that appear in our model. Panels C-E further elaborate on these dynamics by presenting time series and phase space trajectories (including nullclines) for oscillatory, excitable, and non-excitable states, respectively.}
    \label{fig:Fig1}
\end{figure*}

\section{Model system}\label{sec:sectionII}

Figure \ref{fig:Fig1}A depicts the spatial configuration of the system under investigation, comprising a two-dimensional, disc-shaped structure with radius $R=50$, termed the internal system, surrounded by a one-dimensional ring, referred to as the external system. Both the internal and external systems are modeled using the FHN equation (\ref{EqFHN_ODE}), with the former set in a two-dimensional space and the latter on a one-dimensional ring geometry.\\

\textbf{Internal FHN System:}\\

The internal system's dynamics are governed by:\\
\begin{eqnarray}
        \frac{1}{\tau}\frac{\partial u_i}{\partial t}&=&D_i (\frac{\partial^2 u_i}{\partial x^2} + \frac{\partial^2 u_i}{\partial y^2})-{u_i}^3+u_i-v_i, \\[\jot]
        \frac{1}{\tau}\frac{\partial v_i}{\partial t}&=&D_i (\frac{\partial^2 v_i}{\partial x^2} + \frac{\partial^2 v_i}{\partial y^2})+\varepsilon(u_i-b_i(\boldsymbol{x})v_i+a).
\end{eqnarray}

\textbf{External FHN System:}\\

The external system is described by:\\
\begin{eqnarray}
        \frac{\partial u_e}{\partial t}&=&D_e \frac{\partial^2 u_e}{\partial \phi^2} -{u_e}^3+u_e-v_e, \\[\jot]
        \frac{\partial v_e}{\partial t}&=&D_e \frac{\partial^2 v_e}{\partial \phi^2} +\varepsilon(u_e-bv_e+a_e(u_i)). 
\end{eqnarray}
Here, each node on the external ring is unidirectionally linked to the nearest node on the internal system's outer layer via the term $a_e(u_i)$.\\

\begin{figure}
    \centering
    \includegraphics[width=0.49\textwidth]{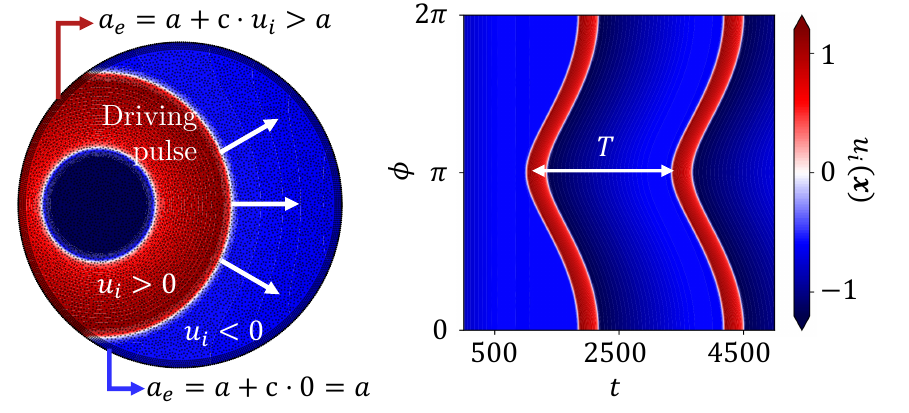}
    \caption{\textbf{Pacemaker-induced traveling pulses.}
    A. Pulses are initiated at the pacemaker region and propagate across the domain as traveling waves. B. Kymograph showing the pulses as they approach and reach the domain's boundary. A new pulse is initiated at intervals of $t=t_0+T$, where $t_0$ marks the initiation of the preceding pulse, ensuring a continuous generation and propagation of pulses to the boundary. See the associated dynamics in Supplementary Movie 1. The simulations are performed with the model's standard parameter set (see Tab. \ref{tab:modelParameters}).}
    \label{fig:Fig3}
\end{figure}

\begin{figure*}
    \centering
    \includegraphics[width=\textwidth]{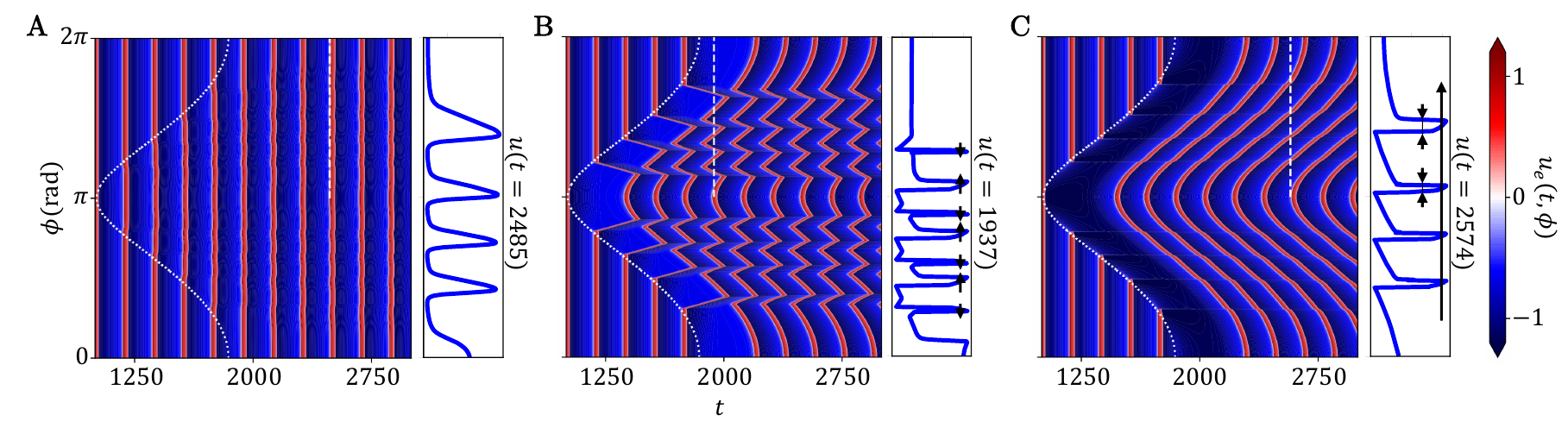}
    \caption{\textbf{External ring dynamics driven by a constant wave pulse.} 
A. In regimes with small coupling, phase patterns emerge (Supp.Movie 2). 
B. At intermediate coupling levels, a coexistence of slow traveling pulses alongside both slow and fast phase waves is observed (Supp.Movie 3). 
C. With large coupling, phase waves exhibiting phase shifts become prominent, characterized by a noticeable contraction in wave profiles (Supp.Movie 4). The simulations are performed with the model's standard parameter set (see Tab. \ref{tab:modelParameters}).
}
    \label{fig:Fig4}
\end{figure*}

Within the internal system, two distinct dynamical regions are demarcated based on parameter choices: region 1 exhibits relaxation oscillations (Fig. \ref{fig:Fig1}B,C), while Region 2 displays excitable dynamics (Fig. \ref{fig:Fig1}B,D).
In the presence of spatial diffusive coupling, this setup allows Region 1 to act as a pacemaker, generating traveling pulses across the domain (Fig. \ref{fig:Fig3}). Note that traveling waves can also emerge in internal Region 2 when it exhibits oscillatory behavior, provided that the pacemaker oscillates faster than its surroundings.   \\

The external system, sharing parameters with Region 1 of the internal system, is inherently oscillatory. However, it features significantly lower diffusion (three orders of magnitude) and a faster time scale (one order of magnitude), reflecting the biological inspiration behind the model.\\

Unidirectional coupling from the internal to the external system is introduced as follows:
\begin{equation}
    a_e(\boldsymbol{x})=\begin{cases}
    a+c\cdot\text{nearest}(u_i) & \text{if }\text{nearest}(u_i)\geq 0\\
    a & \text{otherwise}
    \end{cases},
\end{equation}
where $c$ represents the coupling strength. This coupling mechanism allows the internal system's dynamics to locally influence the external system, potentially driving it towards either excitable (Fig. \ref{fig:Fig1}D) or non-excitable (Fig. \ref{fig:Fig1}E) monostable states, depending on the coupling strength (Fig. \ref{fig:Fig1}B).\\

All the model parameters are given in Table \ref{tab:modelParameters}, and their respective biological motivations are as follows. Both systems exhibit asymmetric spike-like oscillations ($a>0$), with the cytoplasmic system being approximately ten times slower ($\tau$). 
The overall system size (radius R) is set to 50 dimensionless units. For reference, a \textit{Xenopus laevis} frog egg is approximately 1 millimeter in diameter. The pacemaker region (Region 1 - radius r) is chosen to be 10 dimensionless spatial units. Previous studies have demonstrated that the nucleus acts as a pacemaker to initiate waves in the early \textit{Xenopus laevis} cytoplasm \cite{nolet2020nuclei, puls2024mitotic, afanzar2020nucleus}. Both nuclear size and cell size change during the early cleavage stages of the frog embryo, with the nuclear-cytoplasmic ratio ranging from about 0.1\% to 10\% of the total cell size \cite{Pineros2024}. 
The period of our pacemaker region is set to approximately 2200 dimensionless time units, which corresponds to a typical cell cycle period of 30 minutes in the early \textit{Xenopus laevis} embryo. By normalizing time and space in this manner, we derive realistic values for the diffusion coefficients, based on Ref. \cite{burkart2022control}. These coefficients were further fine-tuned to match the observed cytoplasmic and cortical wave speeds \cite{chang2013mitotic, puls2024mitotic, landino2021rho}. Finally, the coupling parameters were chosen to replicate the inhibitory dynamics exerted by the cytoplasm on the cortex.
\\

The internal system functions exclusively as a driver for the external system's dynamics, remaining unaffected by the latter's behavior. Consequently, while the external system can show intricate dynamics stemming from its interaction with the internal system, the internal system itself exhibits quite straightforward behavior that can be precisely defined. Note that changes in the pacemaker position lead to changes in the external dynamics, but changes in the radius do not (Supp.Fig. \ref{supFig:Fig0}).

In this setup, the oscillatory region (region 1), acting as a pacemaker, maintains a consistent oscillation period, resembling the nucleus \textbf{\textcolor{blue}{of $T_i=2200$}}. This rhythmic activity induces excitation in the adjacent region (Region 2), leading to the generation of traveling pulses. Despite the different curvatures, these pulses travel at a constant speed of $v \approx 0.059$ (Supp.Fig. \ref{supFig:Fig1}) and have constant width of $\approx 18$\cite{Tyson1988}. This enables the precise prediction of the pulse's arrival time at the external system's boundary. We thus simplify the simulations by capturing the waveform of the traveling pulse and reintroducing it periodically, aligning with the anticipated timing (Fig. \ref{fig:Fig3}).

\begin{table}[h]
\begin{tabular}{|cc|rl|}
\hline
\multicolumn{2}{|c|}{\textbf{System}}                                                             & \multicolumn{1}{c|}{\textbf{Parameter}} &  \multicolumn{1}{c|}{\textbf{Value}}           \\ \hline
\multicolumn{1}{|c|}{\multirow{11}{*}{\textbf{Circular system}}} & \multirow{6}{*}{\textbf{Internal region}} & $a=$                              & 0.056           \\
\multicolumn{1}{|c|}{}                                  &                                & $b($R1$)=$                    & 0.02            \\
\multicolumn{1}{|c|}{}                                  &                                & $b($R2$)=$                    & 0.10            \\
\multicolumn{1}{|c|}{}                                  &                                & $\varepsilon=$                              & 0.019           \\
\multicolumn{1}{|c|}{}                                  &                                & $D_i=$                              & 0.4              \\
\multicolumn{1}{|c|}{}                                  &                                & $\tau=$                              & 0.086 \\ \cline{2-4} 
\multicolumn{1}{|c|}{}                                  & \multirow{5}{*}{\textbf{External ring}} & $a=$                              & 0.056           \\
\multicolumn{1}{|c|}{}                                  &                                & $b=$                              & 0.02            \\
\multicolumn{1}{|c|}{}                                  &                                & $\varepsilon=$                              & 0.019           \\
\multicolumn{1}{|c|}{}                                  &                                & $D_e=$                              & 0.0001             \\
\multicolumn{1}{|c|}{}                                  &                                & $c=$                              & (0.005,0.1,0.7) \\ \hline
\multicolumn{2}{|c|}{\multirow{6}{*}{\textbf{Linear system: External line}}}                                     & $a=$                              & 0.056           \\
\multicolumn{2}{|c|}{}                                                                   & $b=$                              & 0.02            \\
\multicolumn{2}{|c|}{}                                                                   & $\varepsilon=$                              & 0.019           \\
\multicolumn{2}{|c|}{}                                                                   & $D=$                              & (0.0,0.1,1.0)   \\
\multicolumn{2}{|c|}{}                                                                   & $v_\text{driving}=$                     & 10              \\
\multicolumn{2}{|c|}{}                                                                   & $c=$                              & (0.008,0.1,0.44)    \\ \hline
\end{tabular}

\caption{\textbf{Overview of model parameters.} This table shows the set of model parameters used throughout the simulations, except where specific deviations are noted in individual figure captions. Parameters presented with multiple values correspond to their respective values in the $(\text{Oscillatory},\text{Excitable},\text{Non-Excitable})$ regimes.}
\label{tab:modelParameters}
\end{table}

\section{Three distinctive dynamical regimes in the coupled system}\label{sec:sectionIII}

As the coupling strength $c$ increases, three distinct dynamical behaviors emerge in the external system's outer ring (Fig. \ref{fig:Fig4}). We define the sections of the ring influenced by the internal wave as "driven regions," while areas that remain unaffected are termed "non-driven regions."\\

\textbf{Weak coupling ($0<c<0.017$).}  In this regime, the external system continues its oscillatory behavior even as the internal pulse traverses through it. However, the oscillation period in the driven regions slows down, creating a phase disparity with the non-driven regions. This discrepancy results in "phase patterning" that persists even after the internal pulse exits the external system.\\

\textbf{Intermediate Coupling ($0.017<c<0.295$).} At this level of coupling, the external system transitions to an excitable state in response to the internal pulse. Consequently, when an oscillation from a non-driven region intersects with a driven region, it triggers the latter to generate traveling pulses. These pulses continue periodically even after the internal pulse is gone, effectively becoming phase waves. Additionally, faster phase waves, mirroring the shape of the initial internal pulse, emerge and travel in the opposite direction, leading to interactions and eventual annihilation upon collision.\\

\textbf{Strong Coupling ($c>0.295$).} With strong coupling, the external system becomes non-excitable during the passage of the internal pulse, yet minor perturbations from equilibrium are still possible. Phase waves closely tied to the internal pulse form and traverse the ring, ultimately self-annihilating. Notably, these phase waves tend to widen and contract over time.\\

The delineation of these dynamical regimes is primarily governed by the onset of a Hopf bifurcation at $c=0.017$, marking one boundary, and the transition to excitability around $c=0.295$, setting the other limit. However, it is important to note that for type II excitability, the excitability threshold is not clearly demarcated. Thus while the former threshold can be computed analytically (see Supp.Mat.\ref{Hopf}), the latter has to be computed numericaclly.\\

We have characterized many of the observed patterning and wave behaviors as phase phenomena or pseudowaves, where phase disparities give rise to apparent traveling waves, independent of the system's diffusivity \cite{winfree1980geometry}. These dynamics are not constrained in wave speed, and diffusion tends to dissipate these effects over time. In Supp.Fig. \ref{supFig:Fig2}, we demonstrate how diffusion leads to the vanishing of these induced phenomena, and we show the different decay of the maximum phase gradient for different diffusion coefficients, aligning with the phase phenomena hypothesis. Given a small diffusion, this decay process extends well beyond the intervals of the interior driving waves, which periodically reintroduce phase differences, thus mimicking the persistence of actual traveling waves and interactions akin to Turing/Hopf dynamics. The subsequent sections look into the origins and mechanisms behind these phase-induced behaviors in specific scenarios.

\begin{figure}
    \centering
    \includegraphics[width=0.49\textwidth]{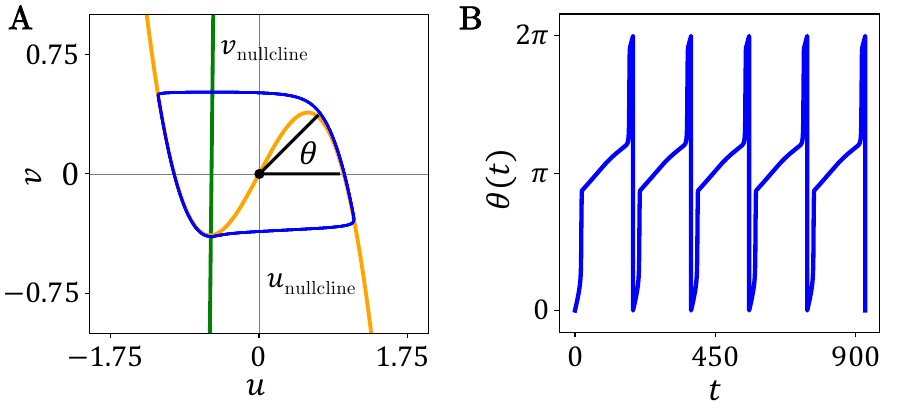}
    \caption{\textbf{From a limit cycle oscillator to a phase oscillator.} A. The depicted limit cycle is transformed into a phase oscillator by employing the arctangent function to define the phase variable. B. This transformation yields a new variable $\theta$ that exhibits periodic oscillations, maintaining the original cycle's two fast and two slow segments. The parameters used here are consistent with those in Fig. \ref{fig:Fig1}C. }
    \label{fig:Fig5}
\end{figure}

\begin{figure*}[t]
    \centering
    \includegraphics[width=\textwidth]{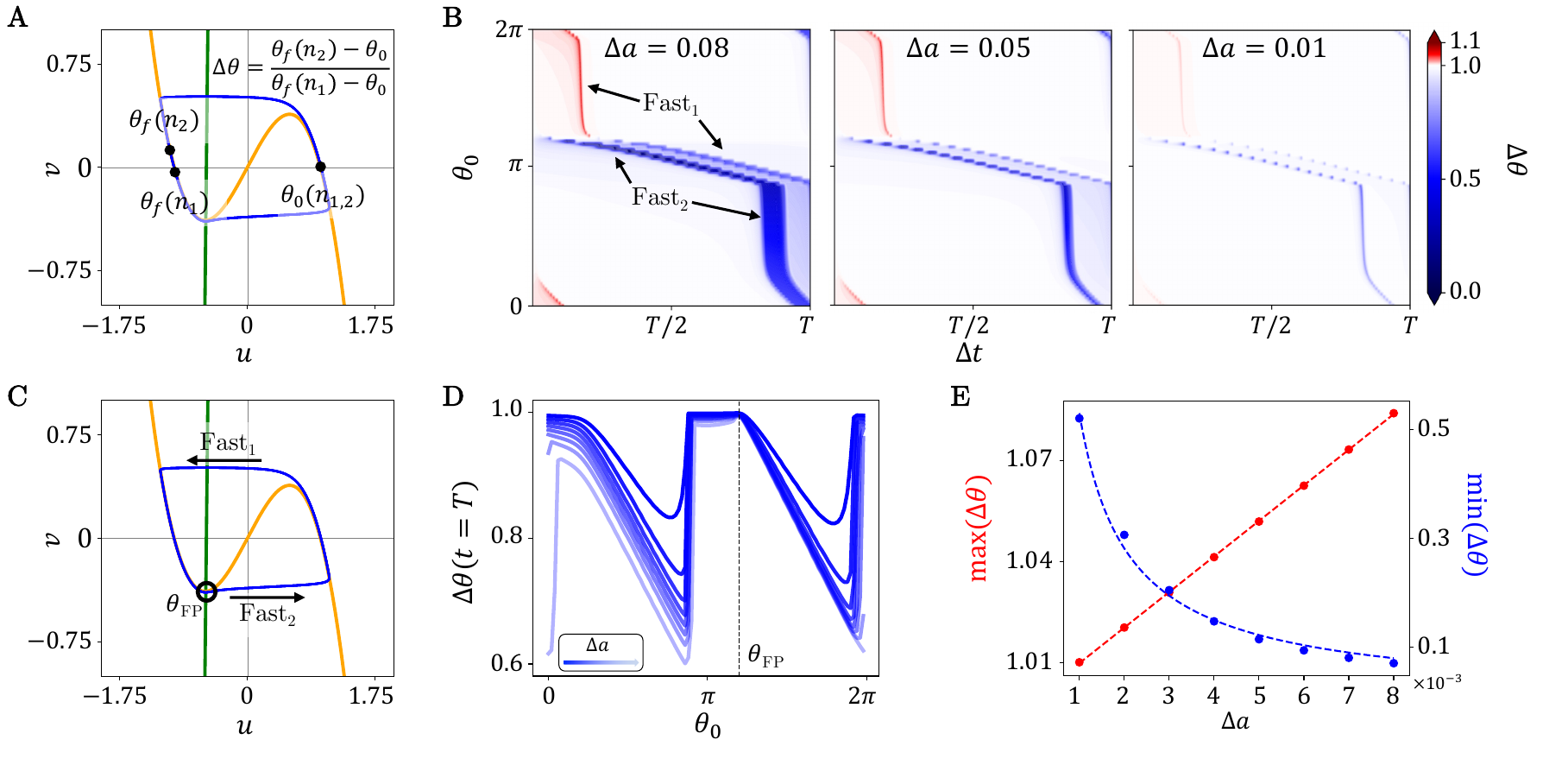}
    \caption{\textbf{Changes in phase velocity when changing system parameters}
A. Conceptual Framework: Illustration of the setup within phase space, highlighting how to quantify changes in phase velocity and accumulated phase changes over time, using the standard parameter set (see Tab. \ref{tab:modelParameters}) while changing the parameter $a$).
B. Dynamics of accumulated phase changes: Visual representation of the phase differences that arise for different initial conditions and durations for three distinct changes of the parameter $a$ (\(\Delta a\)). 
C. Regions in phase space with different velocities, highlighting the "fast" regions as depicted in panel B. 
Close to $\theta_{\text{FP}}$ there is a slow "bottleneck" region.
D. Accumulated phase difference after one complete cycle, in function of various initial phases, and for different \(\Delta a\).
E. The maximum (indicative of faster regions) and minimum (indicative of slower regions) phase differences identified in panel B, across all parameter variations considered in panel D. The maximum phase differences align linearly with modulation intensity (\(\text{max}(\Delta\theta)=(10.538\pm0.001)\Delta a+1.000\)), while the minimum phase differences exhibit an exponential decay relationship (\(\text{min}\Delta\theta=0.001(\Delta a)^{-0.913\pm 0.001}\)).   }
    \label{fig:Fig6}
\end{figure*}

\section{Phase patterning in the oscillatory regime}\label{sec:sectionIV}

In scenarios with small coupling strength, we observe phase patterning in the external system, which remains within the oscillatory region, even though the parameter $a_e$ increases. This phenomenon is analyzed by simplifying the system to a phase oscillator model, as illustrated in Fig. \ref{fig:Fig5}, applicable under conditions of invariant limit cycle shapes, which requires minor parameter adjustments and weak oscillator coupling (see Supp.Fig. \ref{supFig:Fig3}).\\

In the FHN model, the oscillation period lengthens as the parameter $a$ increases, and disappears at the Hopf bifurcation point. Consequently, regions impacted by a driving wave demonstrate slower oscillations, expected to resynchronize after the internal wave has passed. Nonetheless, simulations persistently exhibit phase patterning, which we attribute to the non-uniform velocity across the limit cycle, shaped by the driving wave's duration and intensity.\\

To investigate this, we analyze two distinct uncoupled nodes, $n_{1}$ and $n_{2}$, starting from the same position on the limit cycle ($\theta_0(n_{1,2}) = \theta_0$). Node $n_{1}$ retains the non-driven system's parameters, while node $n_{2}$'s parameter $a$ is incremented by $\Delta a$. The system is simulated over one period $T$ of node $n_{1}$'s cycle, after which we measure 
the final phases $\theta_f(n_{1,2})$ of both nodes, and we calculate the phase shift using (see Fig.\ \ref{fig:Fig6}A):\\
\begin{equation}
    \Delta \theta = \dfrac{\theta_f(n_2)-\theta_0}{\theta_f(n_1)-\theta_{0}}.
\end{equation}

This method allows for a detailed analysis of the phase shift's extent and distribution (Fig.\ \ref{fig:Fig6}A-B). With larger $a$ values, oscillations become increasingly sharp, altering the phase shift in the cycle's quicker phases for positive $v$ (Fig.\ \ref{fig:Fig6}C). Node $n_{2}$ goes faster through previously slower sections, potentially reaching the rapid phase earlier or reducing the lag with node $n_{1}$, thereby modifying $\Delta \theta$. Nevertheless, an extension in the low $u$ phase results in a delay when entering the rapid, low $v$ phase. Irrespective of the starting conditions, the cycle consistently experiences a slowdown (Fig.\ \ref{fig:Fig6}D-E).\\

When the system is subjected to a driving force for less than a complete oscillation cycle, the resulting phase differences vary significantly based on the initial phase when the driving wave is applied (Fig.\ \ref{fig:Fig6}B). This variation leads to distinct final phase differences across the system, giving rise to phase patterning. 
The most pronounced patterning occurs in areas where the system's bottleneck—the region nearing a stable fixed point ($\theta_\text{FP}$, Fig.\ \ref{fig:Fig6}C) and hence a future attractor—is more frequently influenced by the driving pulse (once, in scenarios where the driving duration is shorter than the oscillation period). The term 'bottleneck' here signifies the critical transition zone in the system. The extent and amplitude of these patterns are further shaped by factors like the driving pulse's width and the variance in system parameters. The resulting phase differences occur periodically in space. Given the driving pulse's constant velocity $v$, the characteristic wavelength of the observed patterning is given by $\lambda = vT$. \\

\begin{figure*}[t]
    \centering
    \includegraphics[width=\textwidth]{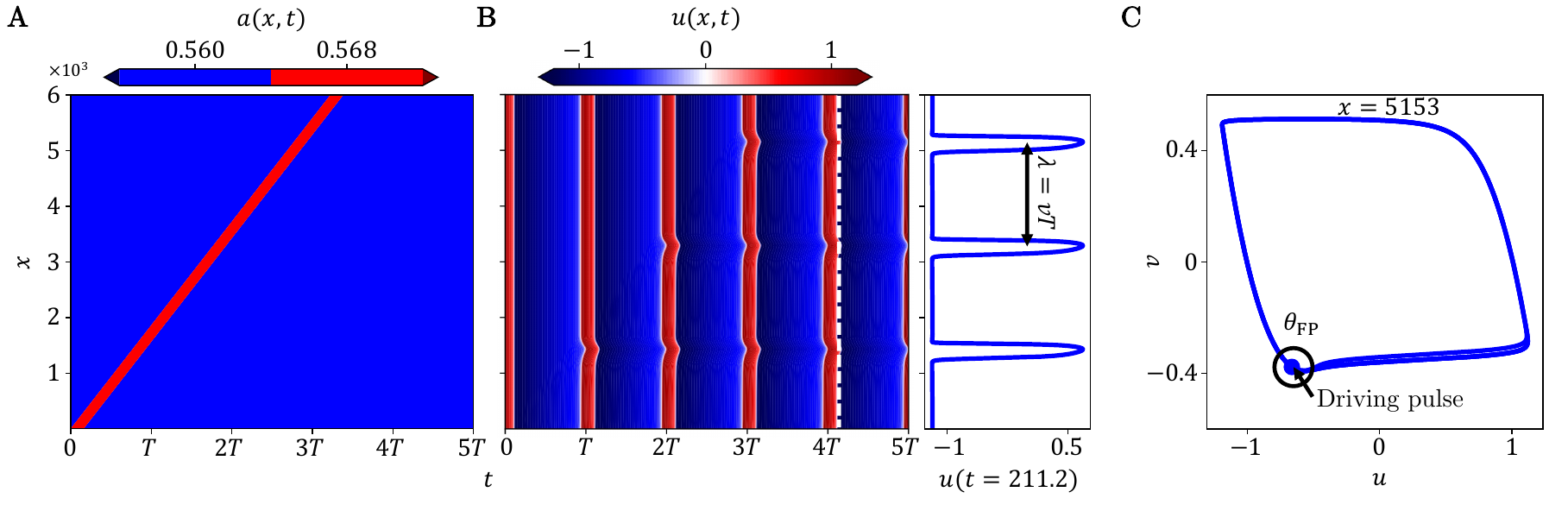}
    \caption{\textbf{Effect of a driving wave on a spatially uncoupled system of FHN oscillators.} 
A. Shows the values of $a$ according to position and time; the driving pulse of constant width and velocity travels throughout the whole system. 
B. Showcases the formation of phase patterning due to the driving wave and illustrates the profile at $t=211.2$. The distance between peaks (patterning wavelength) matches $\lambda=vT$. 
C. Depicts the trajectory of position $x=5153$ and shows the timing at which the driving wave reaches it. The position is within the bottleneck region, and accordingly, it is one of the peaks in the patterning.
All simulations are performed with the model's standard parameter set (see Tab. \ref{tab:modelParameters}). In addition, the inhibitory pulse's width is $10$ units.
}
    \label{fig:Fig7}
\end{figure*}

In Fig. \ref{fig:Fig7}, a simulation within a one-dimensional linear system illustrates the effect of a driving traveling pulse (panel A) - characterized by a constant velocity and width, and a heaviside-like profile - on an oscillatory system with a period of $T\approx186$. The interaction results in phase patterning, as evidenced by a wavelength ($\lambda\approx1860$) (panel B). Notably, the regions encountering the driving wave near the critical bottleneck zone exhibit pronounced phase differences, aligning with theoretical expectations (Panel C).
When this principle is extended to a circular domain, the resulting phase patterning becomes more complex, primarily because the speed of the driving wave observed at the outer ring is no longer constant. Furthermore, the asymmetric profile of the driving pulse in the circular system, in contrast with the symmetric profile in the linear system, results in symmetric (Fig. \ref{fig:Fig7}A) versus asymmetric (Fig. \ref{fig:Fig4}B) peak shapes, respectively.

\begin{figure}[t]
    \centering
    \includegraphics[width=0.49\textwidth]{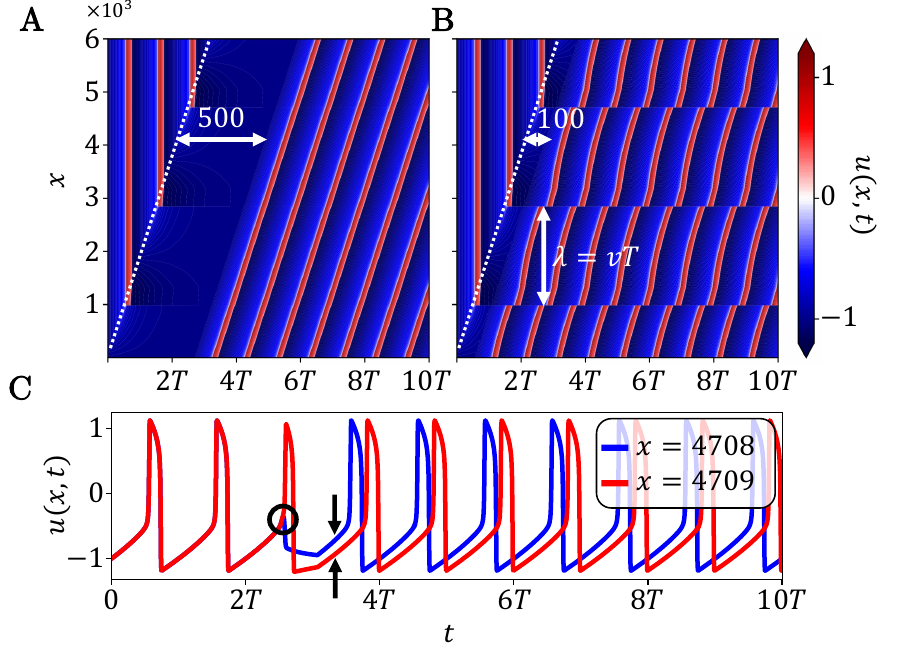}
    \caption{\textbf{Phase waves in the non-excitable regime.} A-B. A constant velocity driving pulse in a 1D linear system with two different durations (A: 500 units, B: 100 units) passes through the domain at a constant velocity. An increase in the pulse's width corresponds to a reduction in phase asynchronies, with predictable intervals at every $\lambda=vT$. C. Time series corresponding to panel B. The phase shift between adjacent nodes (highlighted by the black circle) results in divergent recovery trajectories and durations. The simulations are performed with the model's standard parameter set (see Tab. \ref{tab:modelParameters}). }
    \label{fig:Fig8}
\end{figure}

\section{Phase wave dynamics in the monostable regimes}\label{sec:sectionV}

In this section, we explore the mechanism that governs the formation of phase waves under conditions of intermediate coupling (excitable regime) and strong coupling (non-excitable regime). While both scenarios exhibit a common process for phase wave generation, the excitable regime is distinct for its additional feature of slow-traveling pulses.\\

The previously mentioned $\theta_\text{FP}$ now serves as a stable fixed point, with the earlier bottleneck region becoming part of its basin of attraction. However, the formation of phase waves still fundamentally relies on phase differences.\\

\subsection{Phase waves in the non-excitable regime}
We first examine phase waves that are closely associated with the driving pulse. In a scenario where the driving duration allows the trajectory ample time to reach the fixed point $\theta_\text{FP}$, regardless of the initial phase, all trajectories will converge to this stable fixed point. Following the driving wave's transit, the system transitions back to its oscillatory state. This transition is marked by a re-initiation of oscillations that seem to trail the driving pulse, creating an apparent traveling wave (Fig. \ref{fig:Fig8}A). This effect, however, is actually a manifestation of a phase wave, similar to the dynamics observed in the propagation of an excitable pulse where the system returns to its baseline state \cite{Tyson1988}.\\

What if the system does not have sufficient recovery time? The system's ability to revert to its resting state then hinges on the phase immediately preceding the driving pulse. If the system had just entered the fast region 1 (Fig. \ref{fig:Fig6}C), it would require a larger trajectory to reach the fixed point. On the other hand, if it was about to enter the fast region, the path to $\theta_\text{FP}$ is much shorter. Upon reactivation, the former scenario leads to a significant distance from the fixed point, while in the latter, the system is near the fixed point close to fast region 1, inducing a notable phase shift (Fig. \ref{fig:Fig8}B-C). These shifts play a crucial role in the phase wave's expansion and contraction as depicted in Fig. \ref{fig:Fig4}. The spacing between these shifts is again determined by $\lambda=vT$.\\

\begin{figure}[t]
    \centering
    \includegraphics[width=0.49\textwidth]{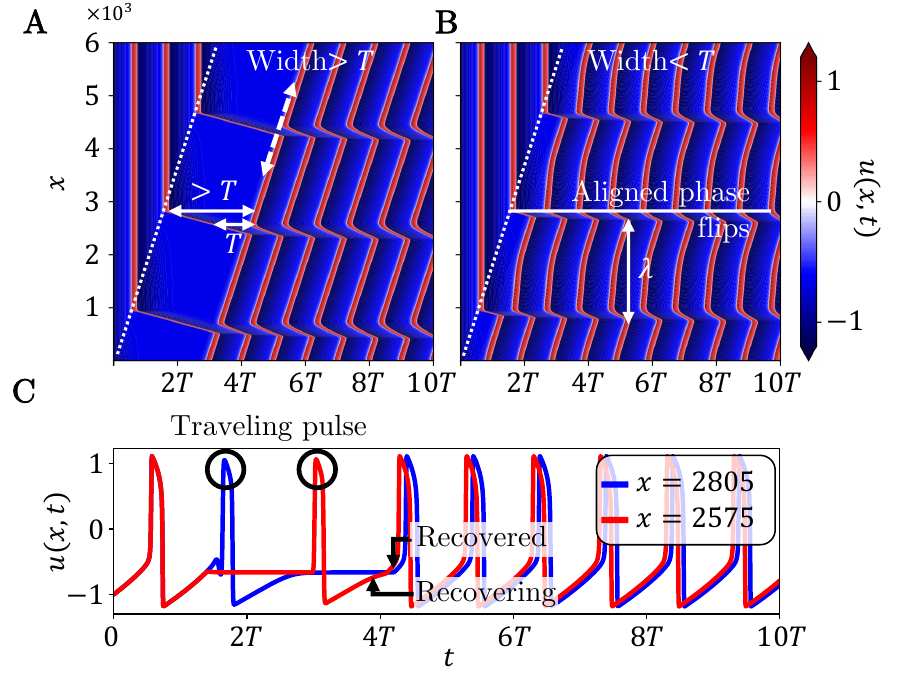}
    \caption{\textbf{Traveling pulses and phase waves in the excitable regime.} A-B. A constant velocity driving pulse in a 1D linear system with two different durations (A: 500 units, B: 100 units) passes through the domain at a constant velocity. Traveling pulses and phase waves propagate in the opposite direction, leading to their eventual annihilation. C. Time series corresponding to panel B, showing the effect of variance in node recovery times. The simulations are performed with the model's standard parameter set (see Tab. \ref{tab:modelParameters}). }
    \label{fig:Fig9}
\end{figure}

\subsection{Traveling pulses and phase waves in the excitable regime}

Comparable to the previous case, phase waves that are synchronized with the driving pulse emerge within the system (Fig. \ref{fig:Fig9}A). In addition, the system becomes excitable under the influence of the driving pulse and is subject to perturbations from adjacent oscillatory regions. These perturbations trigger an extended response that moves across the excitable region as a traveling pulse. These traveling pulses and phase waves move in opposing directions, leading to their mutual annihilation.\\

Even amidst the disruptions caused by the slower-moving traveling pulses, the phase waves linked to the driving pulse persist in their trajectory, as highlighted by the white dashed line. After the passage of the driving pulse, the traveling pulses evolve into phase waves, marked by a distinctive phase flip at the leading edge of the pulse —one oscillatory period after the driving wave passed. The interval between these phase flips, denoted by $\lambda=vT$, resembles dynamics similar to pulse interactions and pacemaker functions, as explored by Michael Stich in 2009 \cite{Stich2009}.\\

However, should the driving pulse's duration be short, it might not allow the system adequate time to converge to its resting state, leading to phase flips that may not coincide with the driving pulse's, as shown in Fig. \ref{fig:Fig9}B-C.

\begin{figure}[t]
    \centering
    \includegraphics[width=0.49\textwidth]{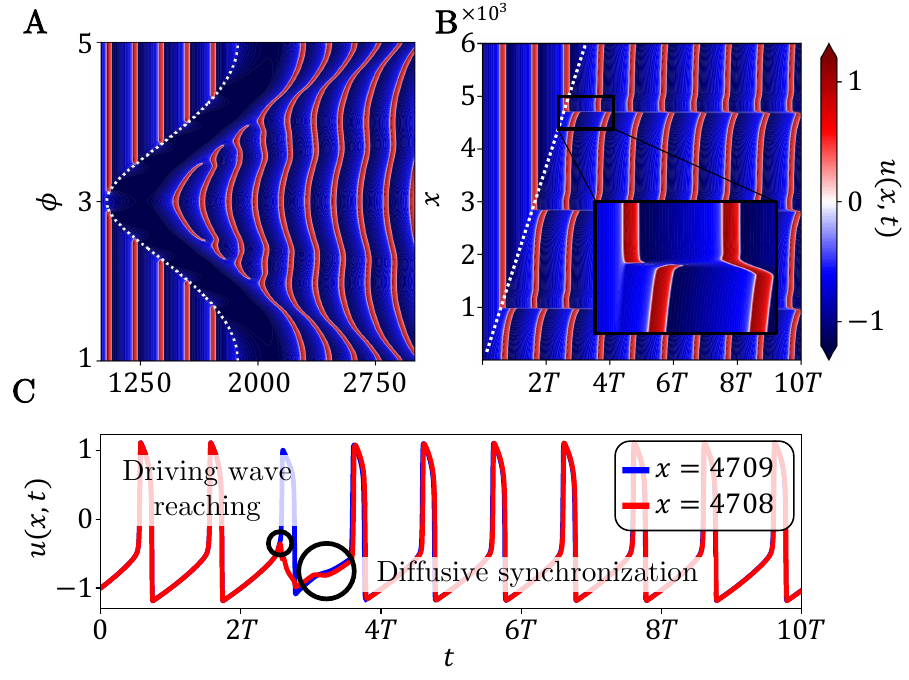}
    \caption{ \textbf{Effects of the diffusion strength.} 
A-B. A constant velocity driving pulse in a 1D circular (A) and linear(B) system passes through the domain at $v=10$. The interplay of a driving pulse and increased diffusion leads to pulses that traverse in the direction opposite to the driving wave. C. Time series corresponding to panel B, showing how the increased diffusion or decreased pulse width lead to a faster synchronization that yields to a regime similar to the excitable case but where the connection is not a reminiscence of a former travelling pulse. The simulations are performed with the model's standard parameter set (see Tab. \ref{tab:modelParameters}), but with $D_e=0.05$. In addition the driving pulse for the linear system has a width of $10$ units.}
    \label{fig:Fig10}
\end{figure}

\section{Effects of the diffusion strength}\label{sec:sectionVI}

As previously noted and shown in Supp.Fig. \ref{supFig:Fig2}, phase waves observed in our system vanish with an increase in the diffusion coefficient, leading to complete synchronization.\\

However, an intriguing phenomenon emerges within the non-excitable regime upon increasing diffusion: a narrow driving wave can cause a significant phase shift. When the affected region begins its oscillation, the diffusion strength is sufficient to influence the neighboring region, even if it has just oscillated. This leads to what we term "connective pulses", where the excitation travels in the direction opposite to the driving wave, as depicted in Fig. \ref{fig:Fig10}. The occurrence of this regime hinges on both the diffusion strength and the driving pulse width, with varying combinations yielding distinct outcomes:\\

In one scenario, larger diffusion lowers the excitability threshold, enabling previously insufficient perturbations to trigger oscillations (Supp.Fig. \ref{supFig:Fig4}A).
Conversely, decreasing the diffusion strength makes the propagation of traveling pulses more challenging, potentially leading to scenarios similar to those illustrated in Fig. \ref{fig:Fig8}, where systems behave as if uncoupled.\\

In another instance, narrowing the driving pulse strengthens the phase shifts. With optimal diffusion levels, the system can link with the nearest excited region, generating connective pulses. Specifically, phase shifts greater than $\pi$ prompt the system to connect with the forthcoming excited region, producing connective pulses that travel opposite to the driving wave (Fig. \ref{fig:Fig10}). On the other hand, phase shifts smaller than $\pi$ lead to connections with the preceding excited region, resulting in connective pulses that propagate in the same direction as the driving wave (Supp. Fig. \ref{supFig:Fig4}B).

\section{Discussion}\label{sec:sectionVII}

The simplified model presented here has successfully captured a broad spectrum of phase phenomena, which are closely associated with cellular processes \cite{di2022waves, beta2017intracellular, puls2024mitotic}. Here, we explore the connection between specific biological occurrences and each identified regime within the model. Additionally, we will extend the discussion to the potential applicability of these phenomena in various fields, situating the newly identified phase phenomena within the context of existing literature.\\

The non-excitable regime, representing the simplest dynamics, aligns with surface contraction/relaxation waves observed in cells preparing for division \cite{rankin1997surface,hara1980cytoplasmic,anderson2017desynchronizing}. Previous studies have documented waves in the cortex influenced by cytoplasmic Cyclin B - Cdk1 activity levels \cite{bischof2017cdk1,wigbers2021hierarchy}. Viewing these surface contraction waves as phase waves explains the similarity in their propagation speeds with cytoplasmic, rather than cortical, dynamics.\\

In the excitable regime, not only are fast cortical waves like those mentioned above observed, but also the presence of cortical waves moving at varying speeds \cite{wu2018membrane}. While existing research attributes these speed variations to membrane-mediated shape changes, our model proposes an alternative explanation through phase waves driven by cytoplasmic factors. This hypothesis, suggesting a connection between cortical and cytoplasmic wave velocities, warrants further investigation. The model illustrates the coexistence of waves moving at distinct velocities—some matching the external system, influenced by traveling pulses, and others aligning with the internal system, driven by the driving pulse.
Furthermore, we speculate that in a three-dimensional setting, or a two-dimensional external system, the excitable regime might give rise to more complex dynamics such as spirals or even spiral turbulence, phenomena often described in cortical dynamics research \cite{bement2015activator,whitelam2009transformation}.\\
%\textbf{The following paragraph might require some corrections for accuracy and correctness:}
%In the context of oscillatory dynamics within the cortex of single cells, no direct association has been conclusively established with phase wave patterns. However, drawing parallels from excitable dynamics, we hypothesize that adopting a more comprehensive three-dimensional model and integrating system heterogeneities might result in varied phase patterns and potentially lead to spiral turbulence.

For the oscillatory case, while no direct link has yet been established in the case of single-cell dynamics, we theorize that, similar to the excitable dynamics, transitioning to a more realistic three-dimensional model and incorporating system heterogeneities could also lead to diverse phase patterning and potentially spiral turbulence regimes. 
Considering multicellular dynamics in embryonic development of vertebrates, oscillatory phase wave dynamics are believed to play an important role in somitogenesis \cite{miao2024cellular}. During this process, the embryo's body is segmented into a series of structurally similar units known as somites. These somites are formed through the interaction of oscillatory signals and signaling gradients across the embryo's body, where only a specific phase of the oscillation cycle permits cells to differentiate into somites.\\

Next, we explore the significance of phase waves across various disciplines, illustrating the versatility of the phenomena observed in our system. The occurrence of these regimes in different geometries suggests that the presence of a driving and a driven system, even with minimal diffusion, is sufficient for these behaviors to manifest. This opens the door to replicating these setups in diverse fields. In chemistry, phase waves have been pivotal in understanding the Belousov–Zhabotinsky reaction's shift from triggering mechanisms to phase wave dynamics \cite{reusser1979transition}, and as a distinctive regime in oscillatory heterogeneous systems \cite{ortoleva1973phase}. In nanophotonics, a nuanced approach involves controlling electromagnetic wave phases \cite{chen2018phase}, while in magnetostatics, phase shifts occur as spin waves traverse domain walls \cite{hertel2004domain}. Although the underlying mechanisms driving these phase phenomena may differ, their widespread applicability is evident, and alternative mechanisms might unveil novel applications for these dynamics.\\

Also mathematically such phase phenomena have been explored \cite{winfree1980geometry}. Notably, Tyson and Keener's 1988 analysis of back waves shares parallels with the strong coupling or non-excitable cases observed in our system, where wave fronts propagate via diffusion and the trailing waves emerge from the intrinsic relaxation of excited states, forming a phase wave linked to the advancing front \cite{Tyson1988}.
Similarly, Michael Stich et al.'s investigation into pacemaker interactions with wave trains presents dynamics similar to those in the intermediate coupling or excitable regimes of our study, where phase flips arise with pacemakers replacing the fast phase waves \cite{Stich2009}.
To date, the specific phase patterning phenomenon we describe remains quite unexplored, particularly in its sensitivity to diffusion coefficients, which can rapidly alter or extinguish the pattern as diffusion increases.\\

\section{Conclusions}\label{sec:sectionVIII}

Our simple setup of interconnected FitzHugh-Nagumo (FHN) models, inspired by cellular structures, captures a wide range of phase-related dynamics. In this setup, a traveling wave within the internal system, analogous to a cell's cytoplasm, drives the dynamics of the external system, similar to a cell's cortex, without being influenced in return. The passage of this driving wave pulse through the external system can trigger one of three distinct regimes: oscillatory, excitable, and non-excitable. In the oscillatory regime, phase patterning emerges. The excitable regime is distinguished by the coexistence of phase waves at different velocities and traveling pulses. Meanwhile, the non-excitable regime has phase waves that become distorted.\\

Through careful analysis, we have confirmed the nature of these phenomena and interpreted them in a simplified linear system. This analysis has also established connections between the observed behaviors and characteristics of the driving pulse, such as its width and velocity, thus providing us with the means to manipulate the system to either accentuate or mitigate certain phase phenomena.\\

Looking ahead, our aim is to establish a quantitative relationship between the documented phase phenomena and specific biological occurrences. We also plan to extend our simulations to larger scales to investigate whether these behaviors, particularly when combined with heterogeneities, might give rise to spiral turbulence, a pattern frequently encountered in biological systems. Exploring the consequences beyond biology is another intriguing prospect. Our findings highlight the potential for misinterpreting events, suggesting that what may appear as traveling waves in systems with an underlying driving mechanism and low diffusion could actually be phase wave phenomena.

\section{Ackowledgements}
L.G. acknowledges funding by the KU Leuven Research Fund (grant number C14/23/130) and the Research-Foundation Flanders (FWO, grant number G074321N). A.B.G acknowledges funding from the Biotechnology and Biological Sciences Research Council (BB/W013614/1) and the Leverhulme Trust (RPG-2020-220).

\section{Data availability statement}
The numerical codes to reproduce the figures in this study are openly available in \textsc{GitLab}\cite{GitLab_rep}, and as an archived repository on RDR by KU Leuven\cite{8LYPUH_2024}.

\nocite{*}
\bibliography{biblio}% Produces the bibliography via BibTeX.

\appendix
\section{Integration methods and code availability}\label{Ap.A}

In the circular model setup (as illustrated in Fig. \ref{fig:Fig3} and Fig. \ref{fig:Fig4}), the Euler method, with an order of accuracy of $O(h^3)$, is employed for temporal integration. The spatial component is discretized using the pseudospectral Radial Basis Functions - Finite Difference approach, a mesh-free technique well-suited for this particular study. This method approximates the Laplacian by considering a preset number of neighboring points, referred to as the stencil size ($n$), and weights the contribution of each of them depending on their distance, a so-called shape parameter ($\epsilon$) and a functional form (see Tab. \ref{tab:integrationParameters}). \\

The interior system features a configuration of nodes that are equidistantly ($h_x(\text{normal})$) and randomly placed, along with five densely packed layers of nodes at predetermined angular coordinates ($h_x(\text{dense})$), a density also applied in Region 1. This arrangement ensures the algorithm's convergence under Neumann boundary conditions, which are consistently applied throughout this study. The exterior system is designed with a node density four times greater in its outermost layer than in the interior, establishing a coupling ratio where four external nodes correspond to a single node within the disc (see Tab. \ref{tab:integrationParameters}).\\

For the linear one-dimensional models depicted in Fig. \ref{fig:Fig6}, Fig. \ref{fig:Fig7}, Fig. \ref{fig:Fig8}, Fig. \ref{fig:Fig9} and Fig. \ref{fig:Fig10}, temporal integration is carried out using a 4th-order Runge-Kutta method ($O(h^5)$). The spatial domain is discretized via a finite difference technique, employing centered differences for the Laplacian and achieving an order of accuracy of $O(h^3)$ (see Tab. \ref{tab:integrationParameters}).\\

Further details on the integration parameters and methodology are included within the annotated source code.\\

\begin{table}[]
\begin{tabular}{|lc|rl}
\hline
\multicolumn{2}{|c|}{\textbf{System}}                                                             & \multicolumn{1}{c|}{\textbf{Parameter}} & \multicolumn{1}{c|}{\textbf{Value}}       \\ \hline
\multicolumn{1}{|l|}{\multirow{11}{*}{\textbf{Circular System}}} & \multirow{6}{*}{\textbf{Internal region}} & $n=$                              & \multicolumn{1}{l|}{30}          \\
\multicolumn{1}{|l|}{}                                  &                                & $\epsilon=$                              & \multicolumn{1}{l|}{5}           \\
\multicolumn{1}{|l|}{}                                  &                                & $f(d,\epsilon)=$     & \multicolumn{1}{l|}{$\sqrt{d^2+\epsilon^2}$} \\
\multicolumn{1}{|l|}{}                                  &                                & $h_x(\text{normal})=$                         & \multicolumn{1}{l|}{0.8}         \\
\multicolumn{1}{|l|}{}                                  &                                & $h_x(\text{dense}=)$                         & \multicolumn{1}{l|}{0.6}         \\
\multicolumn{1}{|l|}{}                                  &                                & $h_t=$                           & \multicolumn{1}{l|}{0.1}         \\ \cline{2-4} 
\multicolumn{1}{|l|}{}                                  & \multirow{5}{*}{\textbf{External ring}} & $n=$                              & \multicolumn{1}{l|}{7}           \\
\multicolumn{1}{|l|}{}                                  &                                & $\epsilon=$          & \multicolumn{1}{l|}{2}           \\
\multicolumn{1}{|l|}{}                                  &                                & $f(d,\epsilon)=$                         & \multicolumn{1}{l|}{$\sqrt{d^2+\epsilon^2}$} \\
\multicolumn{1}{|l|}{}                                  &                                & $h_x=$                           & \multicolumn{1}{l|}{0.15}        \\
\multicolumn{1}{|l|}{}                                  &                                & $h_t=$                           & \multicolumn{1}{l|}{0.1}         \\ \hline
\multicolumn{2}{|c|}{\multirow{2}{*}{\textbf{Linear system: External line}}}                                     & $h_x=$                           & \multicolumn{1}{l|}{1}                                \\
\multicolumn{2}{|l|}{}                                                                   & $h_t$=       & \multicolumn{1}{l|}{0.01}        \\ \hline
\end{tabular}
\caption{\textbf{Integration parameters used along the paper.} As an exception when $D_e=0.05$ (Fig. \ref{fig:Fig10} and Supp.Fig \ref{supFig:Fig2} A) is considered $h_x$ in the external ring is $0.3$ and $h_t$ in the external and internal systems is $0.2$.}
\label{tab:integrationParameters}
\end{table}

\section{Statistical analysis}\label{Ap.B}

Theoretical functions provide the basis for the analyses presented in Supp.Fig. \ref{supFig:Fig1} and Fig. \ref{fig:Fig6}. For the former, we employ the \texttt{linregress} function from the \texttt{scipy.stats} library, and for the latter, the \texttt{curve\_fit} function from \texttt{scipy.optimize} is utilized (\texttt{scipy} version 1.10.0).

\beginsupplement
\section{Supplementary Material}
\subsection{Hopf bifurcation in the FHN model}\label{Hopf}

The fixed points in the FHN model are determined by the equation without an analytical solution:
\begin{equation}
    bu_0^3+(1-b)u_0+a=0.
\end{equation}

To assess the stability of these numerically derived fixed points, we linearize the equations around them, introducing a perturbation as follows:
\begin{align}
    \left[\begin{pmatrix}
        1-3u_0^2  & -1\\
        \varepsilon & -\varepsilon b 
    \end{pmatrix}-\sigma I_{2\times2}\right]\begin{pmatrix}
        \xi_u\\
        \xi_v
    \end{pmatrix}
    =
    \begin{pmatrix}
        0\\
        0
    \end{pmatrix}.
\end{align}
This results in an eigenvalue problem described by the determinant (Det) and trace (Tr) as:
\begin{equation}
\begin{array}{ll}
    {\rm Det}(J) = \varepsilon(3bu_0^2-b+1),\\[\jot]
    {\rm Tr}(J) = 1-3u_0^2-\varepsilon b.
\end{array}
\end{equation}

The Hopf bifurcation condition, given by Re$[\sigma]=0$ and Im$[\sigma]\neq0$ or equivalently, ${\rm Tr}(J)=0$ and ${\rm Det}(J)>0$, yields the following expressions:
\begin{align}
  u_{\rm H}=\pm\sqrt{\frac{1-\varepsilon b}{3}},  &&  a_{\rm H}=\pm\dfrac{1}{3}(\varepsilon b^2+2b-3)\sqrt{\dfrac{1-\varepsilon b}{3}},
\end{align}
with the bifurcation characterized by the frequency:
\begin{equation}
    \sigma=\pm i\sqrt{{\rm Det}(J)}=\pm i\sqrt{\varepsilon(1-\varepsilon b^2)}. 
\end{equation}

\subsection{Supplementary Figures}
\begin{figure*}[h]
    \centering
    \includegraphics[width=\textwidth]{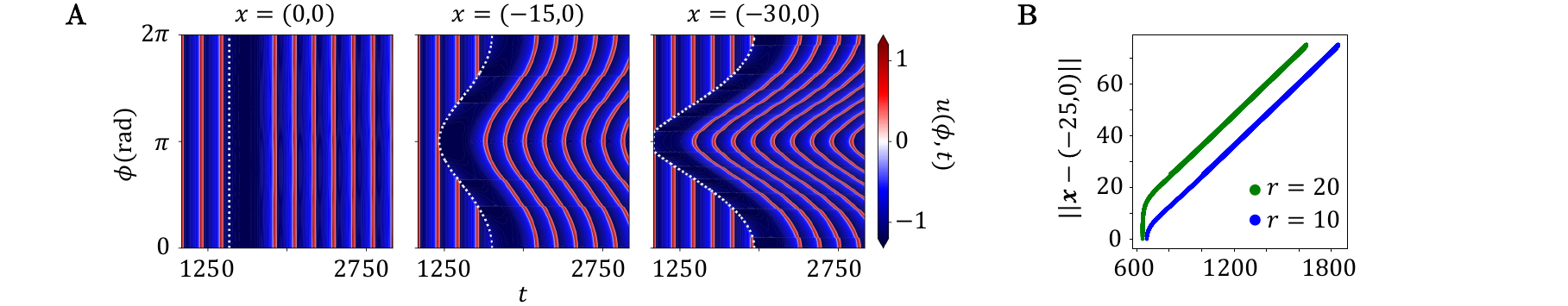}
    \caption{\textbf{Effect of the pacemaker parameters.} A. Changing the position of the pacemaker affects the relative arrival timing to the driven system, thereby modifying the window of emergent dynamics. In the case of a centered pacemaker, no dynamics emerge. B. Changing the radius affects the overall timing of arrival to the driven system but not the relative timing, thus trivially leading to conserved emergent dynamics.}
    \label{supFig:Fig0}
\end{figure*}

\begin{figure*}[h]
    \centering
    \includegraphics[width=\textwidth]{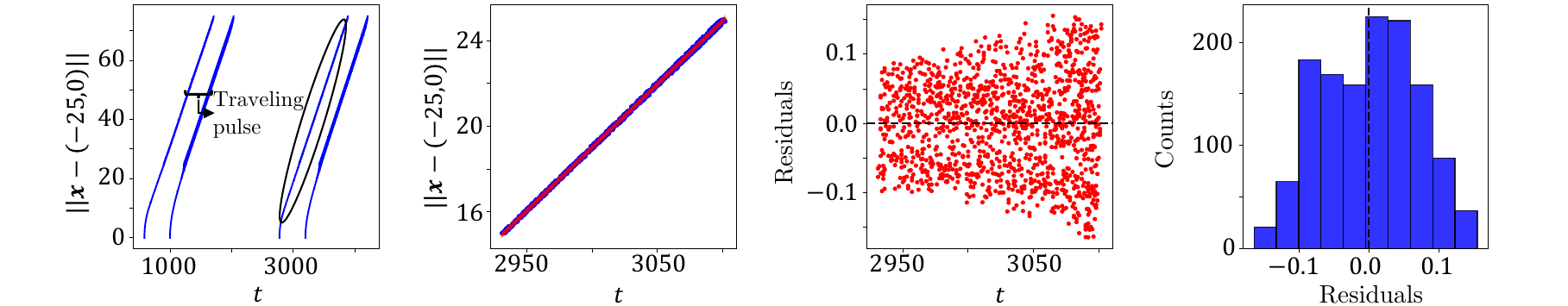}
    \caption{\textbf{Wave velocity of the driving pulses.} The wave fronts are isolated and fitted by a linear regression. The residuals exhibit a centered distribution, ensuring the goodness of fit.}
    \label{supFig:Fig1}
\end{figure*}

\begin{figure*}[t]
    \centering
    \includegraphics[width=\textwidth]{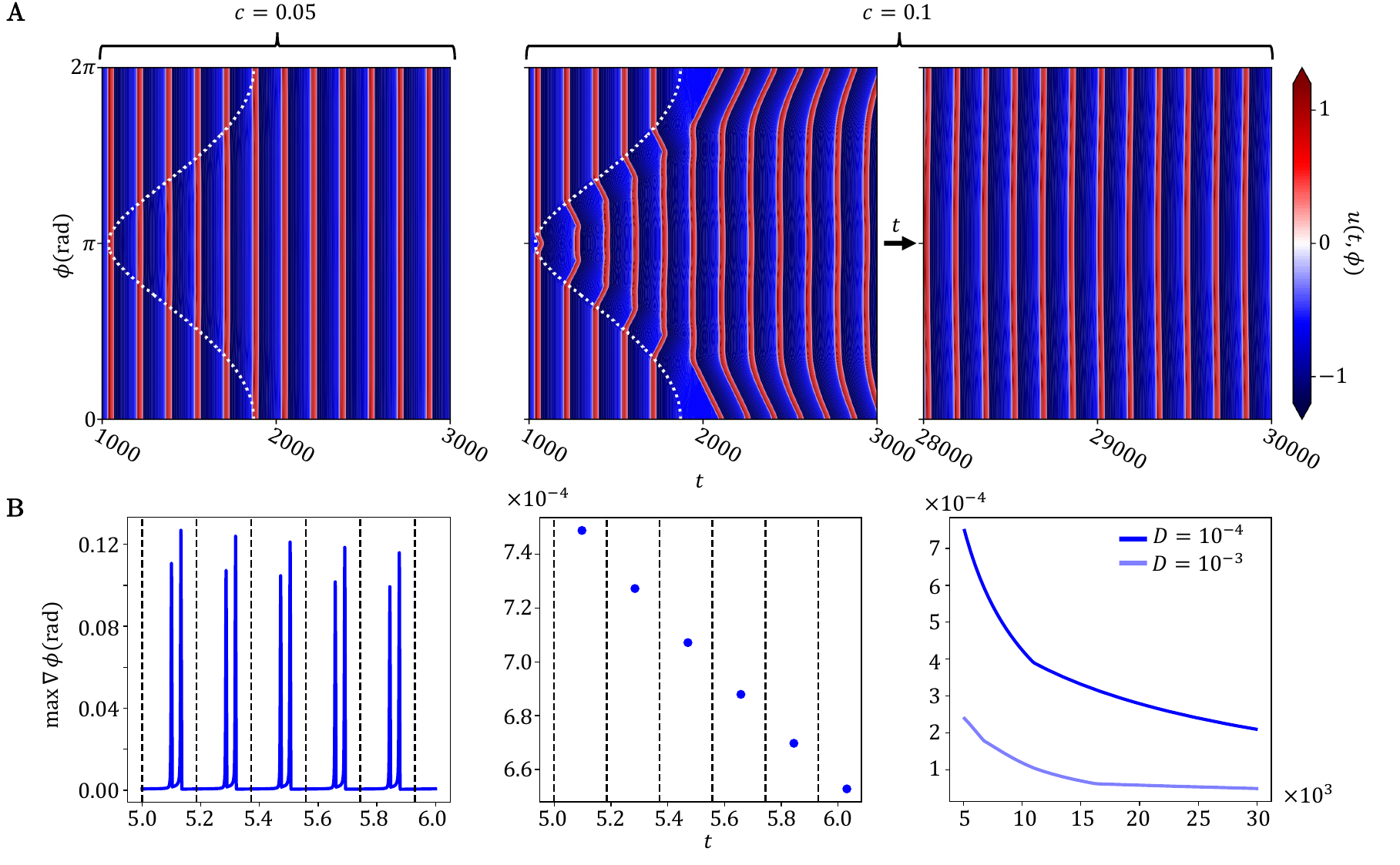}
    \caption{\textbf{Illustrating the phase nature of induced phenomena.} 
A. Shows how by increasing the diffusion coefficient to $D_e=0.05$ and introducing a single driving pulse, the system eventually returns to a synchronized state. This observation confirms that the induced phenomena are related to phase changes rather than the propagation of traveling waves.
B. An alternative approach is used, where the maxima of the phase gradient is monitored over time. Given the system's oscillatory behavior, even as the overall value diminishes, the system continues to exhibit oscillations. By segregating each oscillation and computing the median value, we observe a declining trend in the median values over time. Repeating the same procedure for a larger value of the diffusion coefficient ($D_e=0.001$) shows how the decreasing trend occurs faster for larger diffusion, as expected}
    \label{supFig:Fig2}
\end{figure*}

\begin{figure*}[t]
    \centering
    \includegraphics[width=\textwidth]{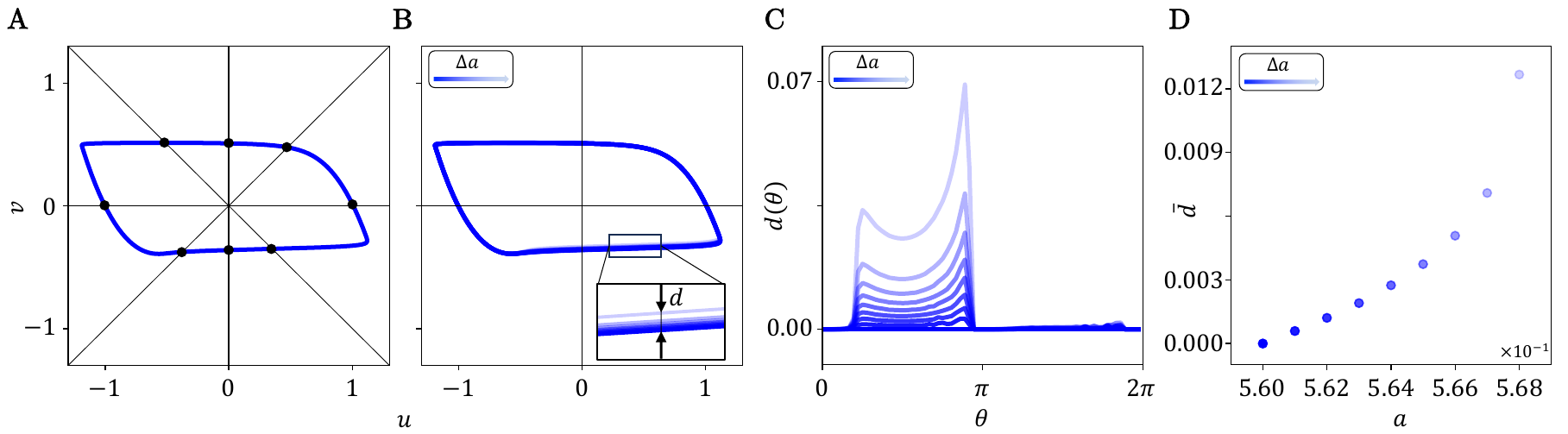}
    \caption{\textbf{Quantifying the changes on the limit cycle with respect to the control variable $a$.}
A. The $(u,v)$ coordinates undergo assessment at every $5^{\circ}$ interval, as depicted in panel. 
B-C. This process involves calculating the distances between the coordinates of each limit cycle and those derived at Region 1 for every angular position, revealing a predictable variation in the distance distribution across the panels. 
D. Subsequently, the average distance for each value of $a$, or equivalently, each $\Delta a$, is determined. When compared to the absolute magnitudes of the variables $(u_\text{max},v_\text{max})=(1.19,0.51)$, it becomes clear that the observed displacement is minimal.}
    \label{supFig:Fig3}
\end{figure*}

\begin{figure*}[t]
    \centering
    \includegraphics[width=\textwidth]{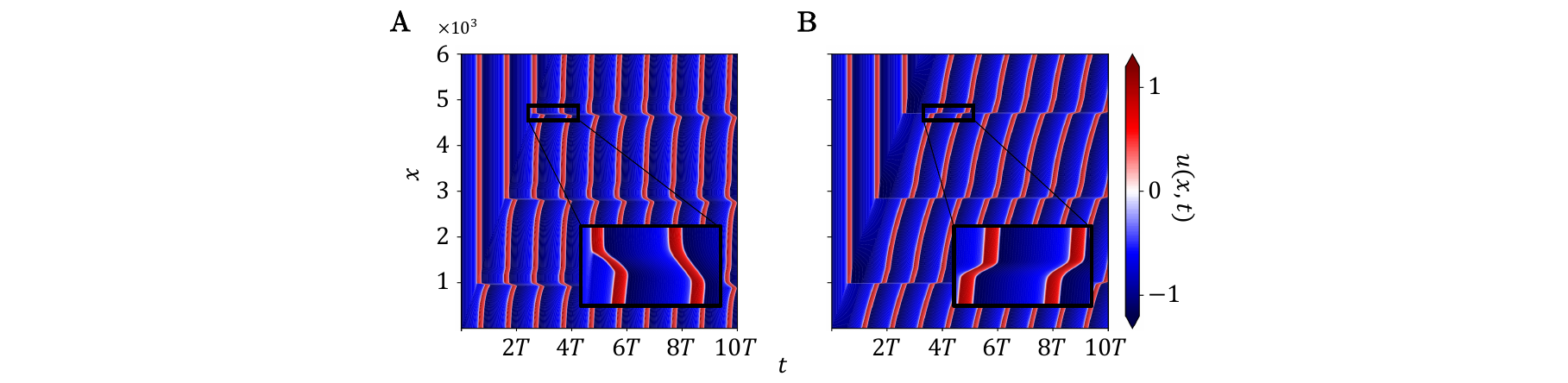}
    \caption{\textbf{Impact of diffusion and pulse width on the non-excitable Regime}. 
A. Illustrates the transition from a monostable regime with opposing pulses to an excitable regime when the diffusion coefficient $D$ is increased from $0.1$ to $0.5$, under the same conditions as depicted in Fig. \ref{fig:Fig10}, with a pulse width of $10$ and velocity $v=10$. 
B. Demonstrates that maintaining $D$ at $0.1$ while expanding the pulse width to $100$ results in diminished phase shifting, leading to connective traveling pulses that align with the driving wave's direction.}
    \label{supFig:Fig4}
\end{figure*}

\end{document}